\documentclass[epj]{svjour}
\usepackage{epsfig}

\newcommand{\avk}{\langle k \rangle} 

\begin{document}

\title{Cut-offs and finite size effects in scale-free networks}

\author{Mari{\'a}n Bogu{\~n}{\'a}\inst{1}
\and Romualdo Pastor-Satorras\inst{2} \and Alessandro
Vespignani\inst{3}}

\institute{Departament de F{\'\i}sica Fonamental, Universitat de
  Barcelona, Avinguda Diagonal 647, 08028 Barcelona, Spain \and
Departament  de F{\'\i}sica i Enginyeria Nuclear, Universitat
Polit{\`e}cnica de Catalunya, Campus Nord, 08034 Barcelona, Spain
\and Laboratoire de Physique Th{\'e}orique (UMR 8627 du CNRS),
B{\^a}timent 210, Universit{\'e} de Paris-Sud, 91405 Orsay Cedex,
France}

\date{Received: date / Revised version: date}

\abstract{We analyze the degree distribution's cut-off
 in finite size scale-free networks.
 We show that the cut-off behavior with the number of vertices $N$
  is ruled by the topological constraints induced by the connectivity
 structure of the network.  Even in the simple case of uncorrelated
  networks, we obtain an expression of the structural cut-off that
  is smaller that the natural cut-off obtained by means of
   extremal theory arguments. The obtained results are explicitly
 applied in the case of the configuration model to recover
the size scaling of tadpoles  and multiple edges.
\PACS{
      {89.75.-k}{} \and
      {87.23.Ge}{} \and
      {05.70.Ln}{}
     }
}

\maketitle

\section{Introduction}

Recent years have witnessed an increasing scientific interest for
the study of complex networks and the dynamical processes taking
place on top of them~\cite{mendesbook,barabasi02}. Indeed, the
complex topological properties shown by many real networks have
large effects on the behavior of several phenomena characterizing
the dynamics and stability of these systems.  These effects are
particularly intriguing in the case of scale-free (SF) networks,
that is, in networks in which the probability $P(k)$ that a vertex
is connected to $k$ other vertices (the degree distribution)
scales as a power law, $P(k) \sim k^{-\gamma}$
\cite{mendesbook,barabasi02}.  In general, \textit{uncorrelated}
SF networks with a degree exponent $\gamma \leq 3$, exhibit the
lack of epidemic and percolation threshold, that can be identified
in terms of critical phenomena with the absence of any critical
point~\cite{pv01a,havlin00,newman00}. These results have been
generalized to several epidemic and percolation
models~\cite{lloydsir,moreno02,newman02b}, even in the presence of
correlations \cite{marian3,vazquez}, and in critical
phenomena~\cite{doro,leone}. In all these systems, the absence of
the critical point finds an explanation in the diverging second
moment $\langle k^2\rangle$ of the degree distribution of SF
networks with $\gamma \leq 3$, that implies unbounded degree
fluctuations in the limit of infinite network size $N\to\infty$.

Real networks, however, have always a finite number of vertices
$N$, and it has been pointed out that finite size effects
reintroduce a positive threshold in critical processes
\cite{lloydsir,pvbrief}. Indeed, the finite size of real networks
introduces a bound in the possible values of the degree, depending
on the system size $N$, which has the effect of restoring a limit
in the degree fluctuations, re-inducing in this way a non-zero
critical point. Therefore, the existence of  bounds for the maximum
degree becomes a relevant element in order to estimate
the critical properties of dynamical systems defined in networks with
SF topologies.

The presence of bounded SF distributions in complex networks has
been observed in several systems \cite{amaral}. In some cases, the
bound or degree cut-off can be explained in terms of a finite
capacity of the vertices to collect connections or due to
incomplete information \cite{amaral,mossa}. In this case, the
value of the cut-off is a constant that depends on the physical
constraints acting on the systems. A second possibility, the one
in which we are interested here, takes place when the cut-off is
purely accounted for by the finite size of the network, as usually
happens in growing networks, that have grown up to a maximum
number of vertices $N$ \cite{moreira,krapivsky}.

In this paper we will reconsider the nature of the cut-offs due to
finite-size effects in SF networks. We will review how to estimate
this cut-off in terms of extreme value theory, and point out how this
estimate is affected when one takes into account the topological
structure of the network. Our considerations will be illustrated by
analyzing an example of uncorrelated network model.

\section{Extreme value theory and the natural cut-off}

The nature of the degree cut-off in finite-size SF networks has
been considered in several instances in network theory. For
example, Aiello {\em et al.} \cite{aiello} proposed to define a
cut-off $k_m$ as the value of the degree for which we expect to
observe at most one vertex, that is
\begin{equation}
  N P(k_m) \sim 1.
\end{equation}
For a SF network, this expression provides a dependence of the
cut-off with $N$ as
\begin{equation}
  k_m(N) \sim N^{1/ \gamma}.
\end{equation}
This definition, however, lacks some mathematical rigor, since it
considers the probability of a single point in a probability
distribution, which is not completely well-de\-fi\-ned in the
continuous $k$ limit for large $N$.

A more physical definition of cut-off was given by Dorogovtsev {\em et
  al.} \cite{dorogorev}, defining it as the value of the degree $k_c$
above which one expects to find at most one vertex,
\begin{equation}
  N \int_{k_c}^\infty P(k) d k \sim 1.\label{cutoff2}
\end{equation}
In this case, we obtain
\begin{equation}
  k_c(N) \sim N^{1/(\gamma-1)},
  \label{eq:2}
\end{equation}
which is known as the \textit{natural} cut-off of the network.

The origin of the natural cut-off, as well as the constraints that
must be imposed to validate its accuracy, can be better understood in
terms of extreme value theory.  Indeed, if we have a random variable
distributed according to the probability density $\rho(x)$ and we draw
$N$ observations of this quantity, $\{ x_i\}$, $i=1 \cdots N$, the maximum
value of this sample, $\max \{ x_i\}$, will, in turn, be a random
variable. Extreme value theory is aimed at finding the statistical
properties of this maximum. In the simplest case, the sample is built
up of independent events and the distribution function, giving the
probability that $\max \{ x_i\}< X$, is simply given by
\begin{equation}
\Pi(X)=\left\{\int^X \rho(y) dy \right\}^{N}.
\end{equation}
The probability density $\pi(X)$ that $\max \{ x_i\}$ is equal to $X$
is therefore the derivative $ d\Pi(X)/dX$ and the cut-off is then
defined as the average value of the extreme value of the sample, that
is,
\begin{equation}
 x_c(N)=\int X \rho(X) N \left\{\int^X \rho(y) dy \right\}^{N-1} dX.
\label{cutoff1}
\end{equation}
This cut-off is always an increasing function of the sample size and
gives us information about the finite size effects of the process
under study. In a scale-free network, the probability density $\rho(x)$
corresponds to the degree distribution, that scales as $P(k)\sim
k^{-\gamma}$.  By substituting $P(k)$ in Eq.~(\ref{cutoff1}), we recover
the natural cut-off scaling given by Eq.~(\ref{eq:2}).  Even though
Eqs.~(\ref{cutoff1}) and (\ref{cutoff2}) give slightly different exact
values for $k_c(N)$, both lead to the correct dependence on the system
size $N$, which is, in most cases, the relevant information.

It is important to recall that this form of the natural cut-off is
obtained under the assumption that all the elements of the sample are
independently drawn from the probability density $P(k)$.  However,
in real networks the degrees of the vertices are not
simply independently drawn from a probability distribution $P(k)$, but
must satisfy some topological constraints due to the network
structure.  Thus, we must include also the structure of the
connections when considering the scaling of the cut-off.

\section{Structural properties of networks and the structural cut-off}

In order to shed some light on this problem we first need to
characterize some structural aspects of networks
\cite{marianproc,hiddenvars}. In what follows we will consider
undirected sparse networks, that is, networks with a well-defined
thermodynamic limit (or, equivalently, constant average degree $\langle k
\rangle$), with $N$ vertices. Let us define $N_k$ as the number of vertices
of degree $k$. This quantity satisfies $\sum_k N_k=N$ which, in the
thermodynamic limit ($N \gg 1$), allows to define the degree
distribution as $P(k)=N_k/N$. The degree distribution $P(k)$ contains
only information about the local properties of vertices, that is, the
number of edges that emanate from each vertex. Thus, we also need to
specify how different degree classes are connected to each other. To
this end, we define the symmetric function $E_{kk'}$, that gives the
number of edges between vertices of degree $k$ and $k'$, for $k\neq k'$,
and two times the number of self-connections\footnote{By
  self-connections we mean connections between vertices in the same
  degree class.} for $k=k'$. This matrix fulfills the identities
\begin{eqnarray}
  \sum_{k'} E_{k k'} &=& k N_k, \\
  \sum_{k, k'} E_{k k'} &=& \avk N = 2 E,
\end{eqnarray}
where $E$ is the total number of edges in the network. This last
identity allows to define, again in the limit $N\gg1$, the
\textit{joint distribution}
\begin{equation}
  P(k,k')=\frac{E_{k k'}}{\langle k \rangle N},
\end{equation}
where the symmetric function $(2-\delta_{k,k'})P(k,k')$ is the
probability that a randomly chosen edge connects two vertices of
degrees $k$ and $k'$. It is easy to see that, in fact, the degree
distribution $P(k)$ can be derived from the joint distribution
$P(k,k')$ as
\begin{equation}
  P(k) = \frac{\avk}{k} \sum_{k'} P(k,k').
  \label{eq:1}
\end{equation}
Therefore, the joint  distribution conveys all the information
at the degree-degree level. In particular, the assortative
(disassortative) character of the correlations in the network, that
is, the tendency of vertices to connect to vertices of the same
(different) degree class, can be quantified by means of the Pearson
coefficient $r$, defined as the correlation coefficient of the joint
 distribution $P(k,k')$ \cite{assortative,newmanmixing}. It is
worth mentioning that all the above quantities are defined outside the
context of any specific model, which make them completely general.

To proceed further, let us define $r_{kk'}$ as the ratio between the
actual number of edges between vertices of degrees $k$ and $k'$,
$E_{kk'}$, and the maximum value for this quantity, $m_{kk'}$.
Assuming that multiple edges are not allowed in the network the
maximum number of edges between two degree classes is\footnote{While
  the restriction of not having multiple edges may appear unnecessary
  under a mathematical point of view, it is usually observed in real
  networks where redundant edges are not considered as a part of the
  network. If multiple edges are instead allowed $m_{kk'}$ is simply
  given by $m_{kk'}=\min\{k N_k,k' N_{k'}\}$.} $m_{kk'} = \min\{k
N_k,k' N_{k'},N_k N_{k'}\}$ and, consequently, the ratio $r_{kk'}$ can
be written as
\begin{equation}
r_{kk'}=\frac{E_{kk'}}{m_{kk'}}=\frac{\avk
P(k,k')}{\min\{kP(k),k'P(k'),NP(k)P(k')\}}. \label{ratio}
\end{equation}
\begin{figure}[t]
  \epsfig{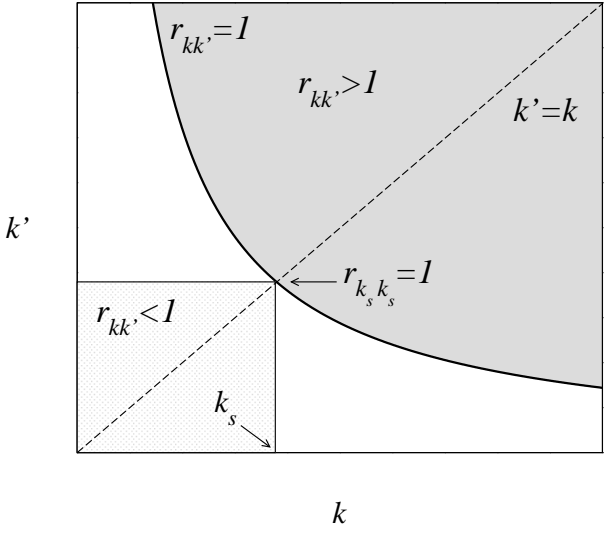}
     \caption{Geometrical construction of the structural cut-off
       $k_s$.}
  \label{diagram}
\end{figure}

A key property of this ratio is that it must be smaller than or equal
to $1$ for any values of $k$ and $k'$, regardless of the type of
network. We can use this simple observation to draw some conclusions
over the value of the cut-off imposed by the structure of the network.
Let us consider, see Fig~\ref{diagram}, the space $k$-$k'$ in which
the joint distribution $P(k,k')$ is defined. The curve $r_{k k'}=1$
defines the boundary\footnote{For simplicity, we have assumed that
  this boundary is given by a smoothly decreasing concave function.
  The same result applies for convex boundaries. More complex
  situations can be considered along the same lines of reasoning.}
separating the region in which the pairs $(k, k')$ take admissible
values ($r_{k k'} \leq 1$) from the unphysical region $r_{k k'} >1$. If
we define a \textit{structural} cut-off $k_s$ as the value of the
degree delimiting the largest square region of admissible values, we
obtain that it is given as the intersection of the curves $r_{ k k'}
=1$ and $k' = k$. That is, the structural cut-off can be defined as
the solution of the implicit equation
\begin{equation}
  r_{k_s k_s} = 1.
\end{equation}
In the following, we will discuss the implications of the structural
cut-off defined in the previous expression.

It is worth noticing that as soon as $k>NP(k')$ and $k'>NP(k)$ the
effects of the restriction on the multiple edges are already being
felt, turning the expression for $r_{kk'}$ to
\begin{equation}
r_{kk'}=\frac{\avk P(k,k')}{NP(k)P(k')}.
\end{equation}
In the case of interest of SF networks these conditions are fulfilled
in the region $k,k'>(\alpha N)^{1/(\gamma+1)}$ (where $\alpha$ is constant
depending on de details of the function $P(k)$), well below the
natural cut-off. As a consequence, this scaling behavior provides a
lower bound for the structural cut-off of the network, in the sense
that, whenever the cut-off of the degree distribution falls below this
limit, the condition $r_{kk'}<1$ is always satisfied.

\subsection{Uncorrelated networks}

Let us first analyze the class of uncorrelated networks. In this case
the joint distribution factorizes as
\begin{equation}
  P_{\rm   nc}(k,k')=\frac{kk'P(k)P(k')}{\avk^2}
\end{equation}
 which, in turn, implies that the
ratio $r_{kk'}$ takes the simple form \cite{hiddenvars}
\begin{equation}
  r_{kk'}=\frac{kk'}{\avk N}. \label{ratio_uncorr}
\end{equation}
In this case, the structural cut-off needed to preserve the
physical condition $r_{kk'}\leq 1$ takes the form
\begin{equation}
  k_s(N) \sim (\avk N)^{1/2},
\end{equation}
independent of the degree distribution, and in particular, of the
degree exponent $\gamma$ in SF networks.  This structural cut-off, which
has been already discussed in the context of the random network model
proposed by Chung and Lu \cite{chunglumode,originnewman}, and also by
Ref.~\cite{krzywickirandom}, coincides with the natural cut-off when
the exponent of the degree distribution is $\gamma=3$ (for instance, the
Barab{\'a}si-Albert network \cite{barab99}). For $\gamma > 3$, the structural
cut-off diverges faster than the natural cut-off, and therefore the
latter should be selected as the appropriate one
\cite{krzywickirandom}.  For $\gamma <3$, however, the exponent of the
natural cut-off is greater than $1/2$ and, as a consequence, the
cut-off predicted by extreme value theory is diverging faster than the
structural one. In other words, this means that uncorrelated SF
networks without multiple edges and exponent $\gamma<3$ must possess a
cut-off that behaves as the structural cut-off and is thus smaller
than the one predicted by the extreme value theory. If this is not the
case, that is, if the actual cut-off is imposed to be larger than the
structural cut-off $k_s$, this means that the network is not totally
uncorrelated and some negative correlations, such as those observed in
the Internet \cite{alexei}, must appear in order to fulfill the
constraint $r_{kk'}\leq 1$~\cite{originnewman,maslovcorr}.

\subsection{Correlated networks}

For correlated networks the position of the cut-off will depend, in
general, on the nature of the correlations through the specific form
of $r_{kk'}$. For assortative networks, that is, networks with
positive degree correlations, the cut-off must be even smaller than
the uncorrelated one since, in these class of networks, high degree
vertices connect preferably to other high degree vertices increasing,
thus, the value of $r_{kk'}$ at $k \simeq k'$. This could explain, for
instance, the appearance of an abrupt cut-off after a power law regime
in many social networks, such as the network of movie actors
\cite{albert00}, which have been found to show assortative mixing. In
the opposite case of disassortative networks the cut-off can be higher
because, in this case, high degree vertices connect preferably to low
degree ones and, as a consequence, the ratio $r_{kk'}$ is reduced near
the region $k \simeq k'$. In summary, the structural cut-off for SF
networks lies somewhere within the interval $[(\alpha N)^{1/(\gamma+1)},(\beta
N)^{1/(\gamma-1)}]$, where $\alpha$ and $\beta$ are characteristic constants,
depending on their correlation structure. In particular, $k_s \sim
N^{1/2}$ corresponds to uncorrelated networks whereas greater values
of the exponent indicate disassortative correlations and smaller
values assortative mixing by degree \cite{moreira,krapivsky}.

\section{The configuration model}

\begin{figure}[t]
  \epsfig{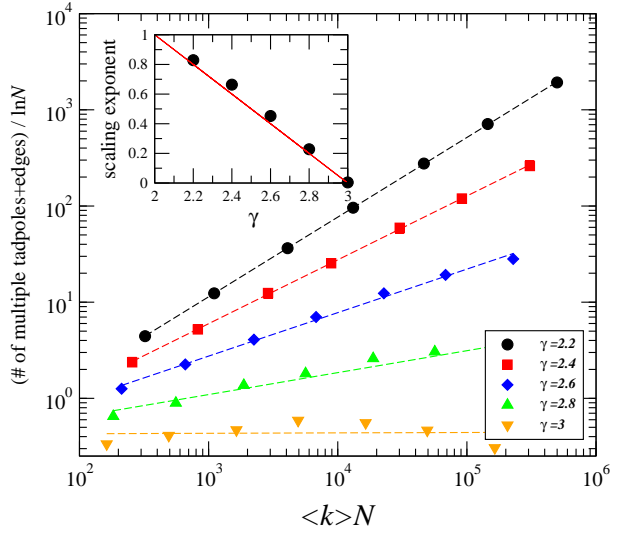}
     \caption{Number of multiple edges plus tadpoles
       satisfying condition (\ref{condition}) as a function of the
       size of the network, $\langle k \rangle N$, for different values of the
       exponent $\gamma$ in the configuration model. The inset shows the
       scaling exponents compared to the theoretical value $3-\gamma$.
       Each point corresponds to an average over $10^4$ networks.  }
  \label{loops}
\end{figure}

In order to check our previous arguments, we shall consider the
configuration model (CM), proposed by Molloy and Reed
\cite{molloy95,molloy98} as a practical algorithm to generate
uncorrelated random networks with a designed degree distribution. The
model first generates a sequence of $N$ degrees, which are
independently drawn from the distribution $P(k)$, and then it proceeds
by connecting pairs of randomly chosen edge ends. At the
same time, however, the model generates a number of multiple edges
and {\it  tadpoles}, that is, edges connected to the same vertex at
both ends. When the degree distribution has a finite second moment
the fraction of multiple edges and tadpoles over the total number
of edges vanishes in the thermodynamic limit and, as a consequence,
they can be neglected. For SF networks with exponent $\gamma <3$, the
situation is different \cite{maslovcorr} as it can be shown by
applying the reasoning concerning the scaling of the degree
distribution cut-off. The problematic degree classes
are those connecting vertices of degrees satisfying
\begin{equation}
  kk'>\langle k \rangle N.
\label{condition}
\end{equation}
For an uncorrelated SF network with degree exponent $2 < \gamma <3$, the
total number of edges satisfying this condition can be calculated, in
the limit $N\gg1$, as
\begin{eqnarray}
  \sum_{kk'>\langle k \rangle N} E_{k k'} & \sim&  N \langle k\rangle \int^{N} d k \int^{\infty}_{\langle k \rangle
    N / k} d k' P_{\rm nc}(k, k') \nonumber \\
  & \sim &  \left(\langle k\rangle N\right)^{3-\gamma} \ln N.
\end{eqnarray}
These edges correspond to an unphysical situation and therefore
must be balanced by a similar scaling of multiple edges plus tadpoles
in order to  restore a physical structure for the network
connectivity. We thus  expect the same scaling law for the number of
multiple edges plus tadpoles satisfying condition (\ref{condition})
generated by the CM algorithm. In order to check this {\it ansatz} we
have performed numerical simulations using the CM for exponents $\gamma <
3$ and we have computed the number of multiples edges plus tadpoles as
a function of the network size $N$. The results, shown in
Fig.~\ref{loops}, are in very good agreement with our {\it ansatz}.

The weight of these multiple edges with respect to the overall
number of edges is small, a fact that could induce us to conclude
that these multiple edges are not important. This is not
completely true, however, since these multiple edges are not
homogeneously distributed among all the degree classes; i.e. they
are present in classes with degree larger than the structural
cut-off $k_s \sim N^{1/2}$. This implies that, in the
thermodynamic limit, all edges connecting vertices with degrees
satisfying condition (\ref{condition}) will always contain a
finite fraction of multiple edges and tadpoles. It is also
interesting to note that by imposing further constraints to avoid
this problem usually generates further correlations. For instance,
by imposing the restriction that only one edge may connect a pair
of vertices will forbid the natural tendency of high degree
vertices to connect mutually and favor their linking with small
degree vertices, originating the presence of effective
disassortative correlations, as discussed in
Refs.~\cite{maslovcorr,originnewman}.

\section{Conclusions}

In the present work we have analyzed the behavior of the maximum
degree in networks with finite size $N$. The natural cut-off scaling
usually considered is not always the appropriate one. The constraints
imposed by the connectivity structure of networks of finite size
generate spontaneous correlations that introduce a structural cut-off
that in some regimes is diverging slower that the natural one and then
determines the actual scaling of the maximum degree. Strikingly, this
phenomenon occurs also in random uncorrelated networks with SF degree
distribution with $\gamma<3$, that are usually used as a first
approximations to represent many real networked structures.  These
results might be particularly relevant in the evaluation of the
resilience to damage and the spreading of infective agents in SF
networks. Indeed, the absence of intrinsic epidemic and damage
thresholds makes these processes dominated by finite size effects.  In
this case, a careful determination of the cut-off behavior with
respect to the network's size is determinant in the calculation of the
effective thresholds which determines the behavior of these dynamical
processes.

\begin{acknowledgement}
  This work has been partially funded by the European Commission - Fet
  Open Project COSIN IST-2001-33555. R.P.-S.  acknowledges financial
  support from the Ministerio de Ciencia y Tecnolog{\'\i}a (Spain) and from
  the Departament d'Universitats, Recerca i Societat de la Informaci{\'o},
  Generalitat de Catalunya (Spain). We also thank an anonymous
  referee for useful comments.
\end{acknowledgement}


\begin{thebibliography}{10}

\bibitem{mendesbook}
S.~N. Dorogovtsev and J.~F.~F. Mendes, {\em Evolution of networks: From
  biological nets to the {I}nternet and {WWW}} (Oxford University Press,
  Oxford, 2003).

\bibitem{barabasi02}
R. Albert and A.-L. Barab{\'a}si, Rev. Mod. Phys. {\bf 74},  47  (2002).

\bibitem{pv01a}
R. Pastor-Satorras and A. Vespignani, Phys. Rev. Lett. {\bf 86},  3200  (2001).

\bibitem{havlin00}
R. Cohen, K. Erez, D. ben Avraham, and S. Havlin, Phys. Rev. Lett. {\bf 85},
  4626  (2000).

\bibitem{newman00}
D.~S. Callaway, M.~E.~J. Newman, S.~H. Strogatz, and D.~J. Watts, Phys. Rev.
  Lett. {\bf 85},  5468  (2000).

\bibitem{lloydsir}
R.~M. May and A.~L. Lloyd, Phys. Rev. E {\bf 64},  066112  (2001).

\bibitem{moreno02}
Y. Moreno, R. Pastor-Satorras, and A. Vespignani, Eur. Phys. J. B {\bf 26},
  521  (2002).

\bibitem{newman02b}
M.~E.~J. Newman, Phys. Rev. E {\bf 64},  016128  (2002).

\bibitem{marian3}
M. Bogu{\~n}{\'a}, R. Pastor-Satorras, and A. Vespignani, Phys. Rev. Lett. {\bf
  90},  028701  (2003).

\bibitem{vazquez}
A. V{\'a}zquez and Y. Moreno, Phys. Rev. E {\bf 67},  0015101  (2003).

\bibitem{doro}
S. N. Dorogovtsev, A.V. Goltsev and J. F. F. Mendes,
Phys. Rev. E {\bf 66},  016104 (2002).

\bibitem{leone}
M. Leone, A. V{\'a}zquez, A. Vespignani and R. Zecchina,
Eur. Phys. J. B {\bf 28},  191  (2002).

\bibitem{pvbrief}
R. Pastor-Satorras and A. Vespignani, Phys. Rev. E {\bf 65},  035108  (2002).

\bibitem{amaral}
L.~A.~N. Amaral, A. Scala, M. Barth{\'e}l{\'e}my, and H.~E. Stanley, Proc.
  Natl. Acad. Sci. USA {\bf 97},  11149  (2000).

\bibitem{mossa}
S. Mossa, M. Barth{\'e}l{\'e}my, H.~E. Stanley, and L.~A.~N. Amaral, Phys. Rev.
  Lett. {\bf 88},  138701  (2002).

\bibitem{moreira}
A.~A. Moreira, J.~S. Andrade, and L.~A.~N. Amaral, Phys. Rev.
  Lett. {\bf 89},  268703  (2002).

\bibitem{krapivsky}
P.~L. Krapivsky and S. Redner, J. Phys. A 35, 9517 (2002).

\bibitem{aiello}
W. Aiello, F. Chung, and L. Lu, Experimental Math. {\bf 10},  53  (2001).

\bibitem{dorogorev}
S.~N. Dorogovtsev and J.~F.~F. Mendes, Adv. Phys. {\bf 51},  1079  (2002).

\bibitem{marianproc}
M. Bogu{\~n}{\'a}, R. Pastor-Satorras, and A. Vespignani,  in {\em Statistical
  Mechanics of Complex Networks}, Vol.~625 of {\em Lecture Notes in Physics},
  edited by R. Pastor-Satorras, J.~M. Rub{\'\i}, and A. D{\'\i}az-Guilera
  (Springer Verlag, Berlin, 2003).

\bibitem{hiddenvars}
M. Bogu{\~n}{\'a} and R. Pastor-Satorras, Phys. Rev. E {\bf 68},  036112
  (2003).

\bibitem{assortative}
M.~E.~J. Newman, Phys. Rev. Lett. {\bf 89},  208701  (2002).

\bibitem{newmanmixing}
M.~E.~J. Newman, Phys. Rev. E {\bf 67},  026126  (2003).

\bibitem{chunglumode}
F. Chung and L. Lu, Annals of Combinatorics {\bf 6},  125  (2002).

\bibitem{originnewman}
J. Park and M.~E.~J. Newman, Phys. Rev. E {\bf 66},  026112  (2003).

\bibitem{krzywickirandom}
Z. Burda and Z. Krzywicki, Phys. Rev. E {\bf 67},  046118  (2003).

\bibitem{barab99}
A.-L. Barab{\'a}si and R. Albert, Science {\bf 286},  509  (1999).

\bibitem{alexei}
R. Pastor-Satorras, A. V{\'a}zquez, and A. Vespignani, Phys. Rev. Lett. {\bf
  87},  258701  (2001).

\bibitem{maslovcorr}
S. Maslov, K. Sneppen, and A. Zaliznyak, Pattern detection in complex networks:
  Correlation profile of the Internet, 2002, e-print cond-mat/0205379.

\bibitem{albert00}
R. Albert and A.-L. Barab{\'a}si, Phys. Rev. Lett. {\bf 85},  5234  (2000).

\bibitem{molloy95}
M. Molloy and B. Reed, Random Struct. Algorithms {\bf 6},  161  (1995).

\bibitem{molloy98}
M. Molloy and B. Reed, Combinatorics, Probab. Comput. {\bf 7},  295  (1998).

\end{thebibliography}
\end{document}